%% file: Cluster_Ellipticity.tex
\documentclass[useAMS,usenatbib]{mn2e}

\input{affiliations.tex}
\usepackage{graphicx,amsmath,amssymb,color}
\usepackage[normalem]{ulem}
\usepackage{mathptmx}
\usepackage{microtype}

\usepackage{hyperref}

\bibpunct{(}{)}{;}{a}{}{,}

\definecolor{joerg}{rgb}{0.7, 0.4, 0.0}
\definecolor{leon}{rgb}{0.0,0.6,0.3}

\renewcommand{\vec}[1]{\mbox{\boldmath$#1$}}
\renewcommand{\d}{\mathrm{d}}

\graphicspath{{figures/}}

\title[Cluster Orientation Bias and its Impact on Weak
Lensing]{Orientation Bias of Optically Selected Galaxy Clusters and
  its Impact on Stacked Weak Lensing Analyses}

\author[J. P. Dietrich et al.]{J\"org
  P. Dietrich\thanks{Email:dietrich@usm.lmu.de}\Munich\afsep\ExcellenceCluster\afsep\AnnArborPhysics,   
  Yuanyuan Zhang\AnnArborPhysics, 
  Jeeseon Song\AnnArborPhysics,
  Christopher P. Davis\AnnArborPhysics\afsep\KIPAC\afsep\Stanford\afsep\SLAC, 
  \newauthor
  Timothy A. McKay\AnnArborPhysics\afsep\AnnArborAstronomy, 
  Leon Baruah\Sussex, 
  Matthew Becker\KIPAC\afsep\Stanford\afsep\SLAC,
  Christophe Benoist\OCA\afsep\LIneA,
  \newauthor
  Michael Busha\KIPAC\afsep\Stanford\afsep\SLAC,
  Luiz A. N. da Costa\LIneA\afsep\ON,
  Jiangang Hao\FNAL,
  Marcio A. G. Maia\LIneA\afsep\ON,
  \newauthor
  Christopher J. Miller\AnnArborAstronomy,
  Ricardo Ogando\LIneA\afsep\ON,
  A. Kathy Romer\Sussex,
  Eduardo Rozo\KIPAC\afsep\SLAC, 
  \newauthor
  Eli Rykoff\KIPAC\afsep\SLAC,  and
  Risa Wechsler\KIPAC\afsep\Stanford\afsep\SLAC\\
  \Munich Universit\"ats-Sternwarte M\"unchen, Scheinerstr. 1, 81679
  M\"unchen, Germany\\
  \ExcellenceCluster Excellence Cluster Universe, 85748 Garching b. M\"unchen, Germany\\
  \AnnArborPhysics Physics Department, University of Michigan, 450
  Church St, Ann Arbor, MI 48109, USA\\
  \KIPAC Kavli Institute for Particle Astrophysics and Cosmology\\
  \Stanford Physics Department, Stanford University, Stanford, CA,
  94305, USA \\
  \SLAC SLAC National Accelerator Laboratory, Menlo Park, CA, 94025\\
  \AnnArborAstronomy Astronomy Department, University of Michigan, 500
  Church St, Ann Arbor, MI 48109, USA\\
  \Sussex Department of Physics and Astronomy, University of Sussex, UK\\
  \OCA Observatoire de la C\^ote d’Azur, UMR 6202 Cassiop\'ee, BP 4229, F-06304 Nice Cedex
  4, France\\
  \LIneA Laborat\'orio Interinstitucional de e-Astronomia - LIneA, Rua Gal. Jos\'e Cristino
  77, Rio de Janeiro, RJ 20921-400, Brazil\\
  \ON Observat\'orio Nacional, R. Gal. Jos\'e Cristino 77, BR Rio de
  Janeiro, RJ 20921-400, Brazil \\
  \FNAL Center for Particle Astrophysics, Fermi National Accelerator
  Laboratory, Batavia, IL 60510, USA
}

\begin{document}

\date{Accepted 2014 June 24. Received 2014 June 18; in original form 2014 May 10}
\pagerange{\pageref{firstpage}--\pageref{lastpage}} \pubyear{2013}
\maketitle
\label{firstpage}

\begin{abstract}
\input{abstract.txt}
\end{abstract}


\section{Introduction}
\label{sec:introduction}

The abundance of galaxy clusters is an important and powerful probe to
constrain cosmological parameters including the Dark Energy equation
of state parameter $w$ \citep*{2011ARA&A..49..409A}. Mass-observable
scaling relations are typically required to translate easily obtained
mass proxies such as optical or X-ray luminosity into cluster masses,
which are needed for cluster cosmology experiments. These scaling
relations must be calibrated via accurate cluster mass measurements.

Weak gravitational lensing is one of the primary methods to measure
the masses of galaxy clusters. Weak lensing masses can be obtained in
a variety of ways. For massive galaxy clusters, the weak lensing
signal is strong enough to be measurable for individual
clusters. Because the cluster mass function declines exponentially
with cluster mass in the mass range of interest
\citep{1974ApJ...187..425P}, most clusters are too small for
individual mass measurements \citep{1999MNRAS.302..821K}. These are
usually binned by a proxy for their mass, such as the optical
luminosity or richness, to increase the signal-to-noise ratio of the
gravitational shear profile \citep[e.g.,][]{2009ApJ...703.2217S}. The
process of averaging cluster properties in bins is commonly called
``stacking''. The masses of these stacked clusters are inferred either
by inverting their observed surface mass density distribution to
three-dimensional density distributions under the assumption of
spherical symmetry \citep{2007ApJ...656...27J} or by fitting spherical
NFW profiles \citep*{1997ApJ...490..493N} taking miscentring and the
halo-halo correlation -- among other components -- into account
\citep{2007arXiv0709.1159J}.

While individual haloes are triaxial, a stacked halo profile should be
spherically symmetric as long as no orientation bias enters the
selection process. Such an unbiased halo selection, however, is
likely impossible for optical cluster finders. Any optical
identification of galaxy clusters relies on finding a significant,
i.e., above a certain threshold, density contrast with respect to the
surrounding field and background population. In the simplest case this
is simply looking for overdensities of galaxies on the sky
\citep{1958ApJS....3..211A}. The near universality of galaxy cluster
density profiles \citep{1997ApJ...490..493N} and luminosity functions
\citep{1976ApJ...203..297S} can be used to enhance the contrast of
objects looking like galaxy clusters with respect to the background
\citep{1996AJ....111..615P}. The fact that the galaxy population in
clusters is dominated by early-type galaxies of very similar colour,
which depends only on redshift, is often used to mitigate projection
effects and provide redshift estimates of galaxy clusters
\citep[e.g.,][]{2000AJ....120.2148G,2007ApJ...660..239K}. Alternatively,
photometric redshift information can be employed to add depth
information to the galaxy distribution on the sky
\citep[e.g.,][]{2010MNRAS.406..673M}. No matter how sophisticated the
optical cluster finding technique is, clusters that are more compact
on the sky will have a higher contrast with the background and will be
easier to discover.

This bias towards objects that are compact on the sky can lead to an
orientation bias in galaxy cluster selection and also bias their
richness estimates. Prolate (cigar-shaped) clusters with their major
axis aligned with the line-of-sight (LOS) are easier to pick out and
look richer than oblate (pancake-shaped) clusters whose minor axis is
aligned with the LOS. If clusters are selected with such a bias, the
average cluster profile in a given richness bin will not be
spherically symmetric but elongated along the LOS.

In this paper we study the effects of orientation bias on the weak
lensing profiles of stacked galaxy clusters and the resulting biases
in mass estimation. In Section~\ref{sec:predictions} we present
analytic predictions for the impact of averaged cluster ellipticity on
mass estimates obtained from a spherical profile inversion. We study
the magnitude of orientation bias and centring errors using mock
observations described in Sect.~\ref{sec:simulations} and a number of
different optical cluster finders in
Section~\ref{sec:cluster-finders}. We discuss our findings
(Section~\ref{sec:results}) and their impact on cluster cosmology
measurements in Section~\ref{sec:discussion}.

Before proceeding we clarify our nomenclature: We call halos those
collapsed objects that are found by spherical overdensity (SO),
friend-of-friend algorithms, or similar halo finders, in dark matter
$N$-body simulations, irrespective of their galaxy content. Galaxy
clusters are objects identified in observations, for the purpose of
this paper in mock optical observations, as potentially corresponding
to a single collapsed dark matter halo. The mappings between haloes and
galaxy clusters are neither one-to-one nor onto. Throughout this paper
we use a Hubble constant of $H_0 = 70$\,km\,s$^{-1}$\,Mpc$^{-1}$.

\section{Elliptical Haloes}
\label{sec:predictions}

As we will be dealing with elliptical haloes throughout this paper, we
first need to define how to parametrise such haloes and what we mean
when we assign a mass to such an object. Different authors employed
different definitions in previous works \citep[e.g.,][CK07
hereinafter]{2002ApJ...574..538J,2007MNRAS.380..149C}\defcitealias{2007MNRAS.380..149C}{CK07}
and it is important to clearly distinguish them from each other,
decide which definition is best used for a certain purpose, and be
able to convert among them. All of these definitions are
generalisations of the spherical NFW profile
\citep{1997ApJ...490..493N}. For example, in the convention of
\citet{2007MNRAS.380..149C}, who follow \citet{2002ApJ...574..538J}
and \citet{2003ApJ...599....7O}, the halo density depends on the
triaxial radius,
\begin{equation}
  \label{eq:1}
  R^2 = \frac{X^2}{a^2} + \frac{Y^2}{b^2} + \frac{Z^2}{c^2}\;\quad (a
  \leq b \leq c = 1)\;,
\end{equation}
where $X$, $Y$, $Z$ are the Cartesian coordinates with respect to the
halo centre, and the virial mass is defined as the mass inside an
ellipsoid with major axis $R_{200}$ inside which the average
overdensity is $200\rho_\mathrm{c}$, where $\rho_\mathrm{c}$ is the
critical density of the Universe. This mass is given by
\begin{equation}
  \label{eq:2}
  M^\mathrm{CK}_{200} = \frac{800\pi}{3} a b R_{200}^3 \rho_\mathrm{c}\;.
\end{equation}

While this convention is well motivated by the ellipsoidal collapse
model \citep*{2001MNRAS.323....1S} and has been used in estimating the
mass calibration bias from fitting spherical NFW models to elliptical
haloes
\citep[e.g.,][]{2009A&A...499..669D,2010A&A...520A..58I,2014MNRAS.439...48A},
it is probably not ideal for our purposes. We want to understand the
impact of orientation bias on cosmological parameter estimates from
large surveys. As many recent cluster cosmological analyses rely on
the spherical overdensity mass function of
\citet{2008ApJ...688..709T}, we should cast the impact of orientation
bias in terms of the same mass convention.

We expect that no cluster finder will have a preferred angular direction on
the sky. The stacked cluster profile should therefore be azimuthally
symmetric around the LOS, which we take to be the $z$-axis of a
cylindrical coordinate system $(r, \theta, z)$. We can then turn a
spherical cluster profile into a biaxial elliptical profile with two
even and one ``odd'' axis by defining an elliptical radius
\begin{equation}
  \label{eq:3}
  \xi^2 = q^2r^2 + z^2\;,\quad q > 0\,,
\end{equation}
where $q$ is the ratio between the ellipsoid axis along the LOS and
the other axes. This definition differs subtly from eq.~(\ref{eq:1})
and the way an ellipsoid is usually defined in that $q$ can be greater
than $1$. While uncommon, this choice has notational advantages later
on. It is also easier to visualise the transformation of an oblate
ellipsoid over a sphere to a prolate ellipsoid as a continuous
stretching along the LOS. In the parametrisation of eq.~(\ref{eq:1})
an oblate ellipsoid has $q = a < 1, b = c = 1$ and a prolate ellipsoid
has $q = 1/a = 1/b > 1, c = 1$. Adopting the convention of
\citetalias{2007MNRAS.380..149C} would lead to awkward case
distinctions later on and requires a rotation of the coordinate system
if one wants the two even axes of a biaxial ellipsoid to be always in
the plane of the sky. In our choice of the definition of the
elliptical radius, the mass definition corresponding to
eq.~(\ref{eq:2}) is
\begin{equation}
  \label{eq:4}
  M_{200}^\mathrm{ell} = \frac{800\pi}{3} q^{-2} \xi_{200}^3 \rho_\mathrm{crit}\;.
\end{equation}

\begin{figure}
  \resizebox{\hsize}{!}{\includegraphics[clip]{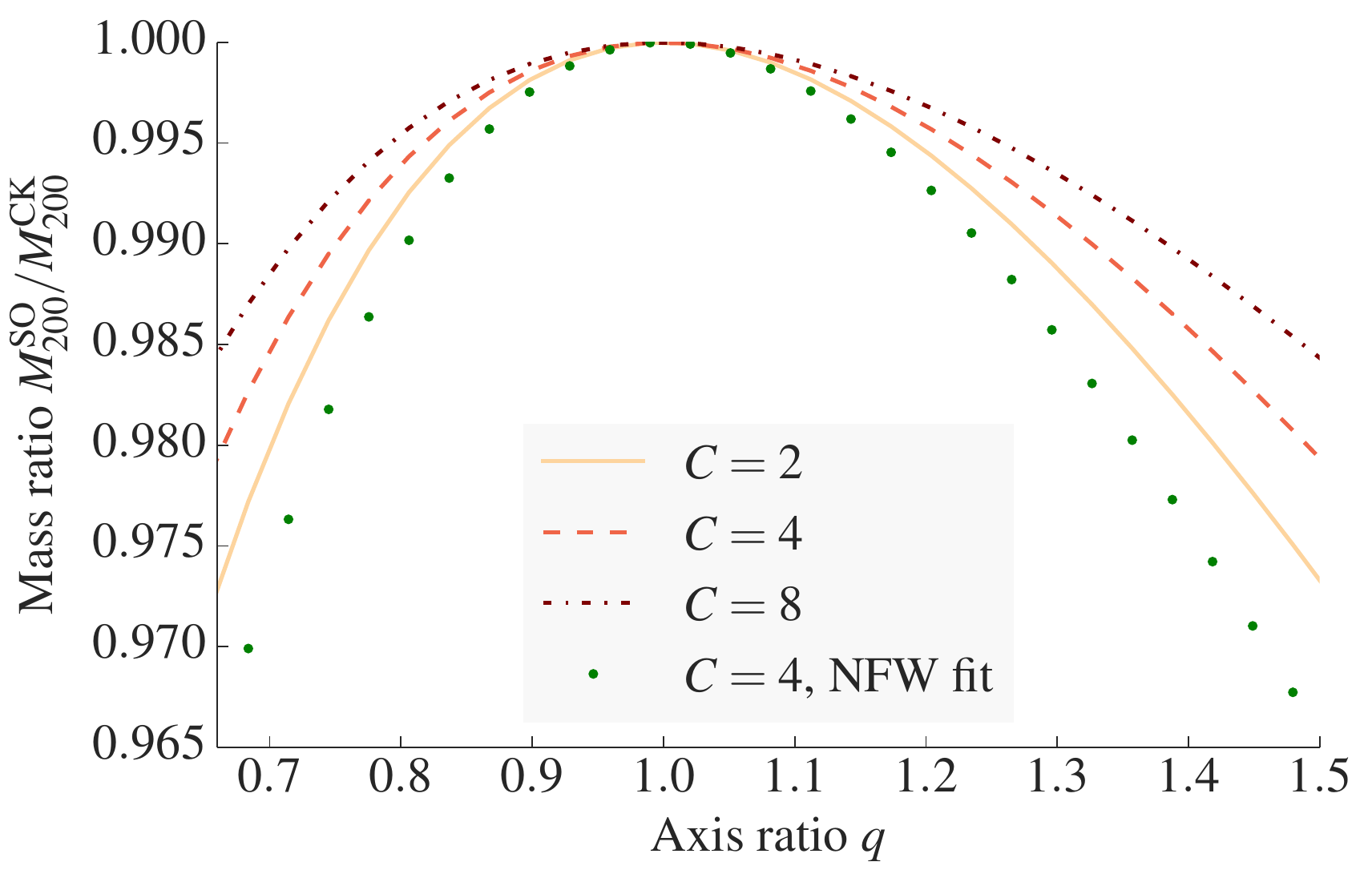}}
  \caption{Conversion between the mass definition of
    \citetalias{2007MNRAS.380..149C} and the spherical overdensity
    mass for different NFW concentrations as a function of axis ratio
    $q$. Also shown is the fit of a spherical NFW profile to the
    average density on spherical shells. }
  \label{fig:CK2NFW}
\end{figure}

\begin{figure}
  \resizebox{\hsize}{!}{\includegraphics[clip]{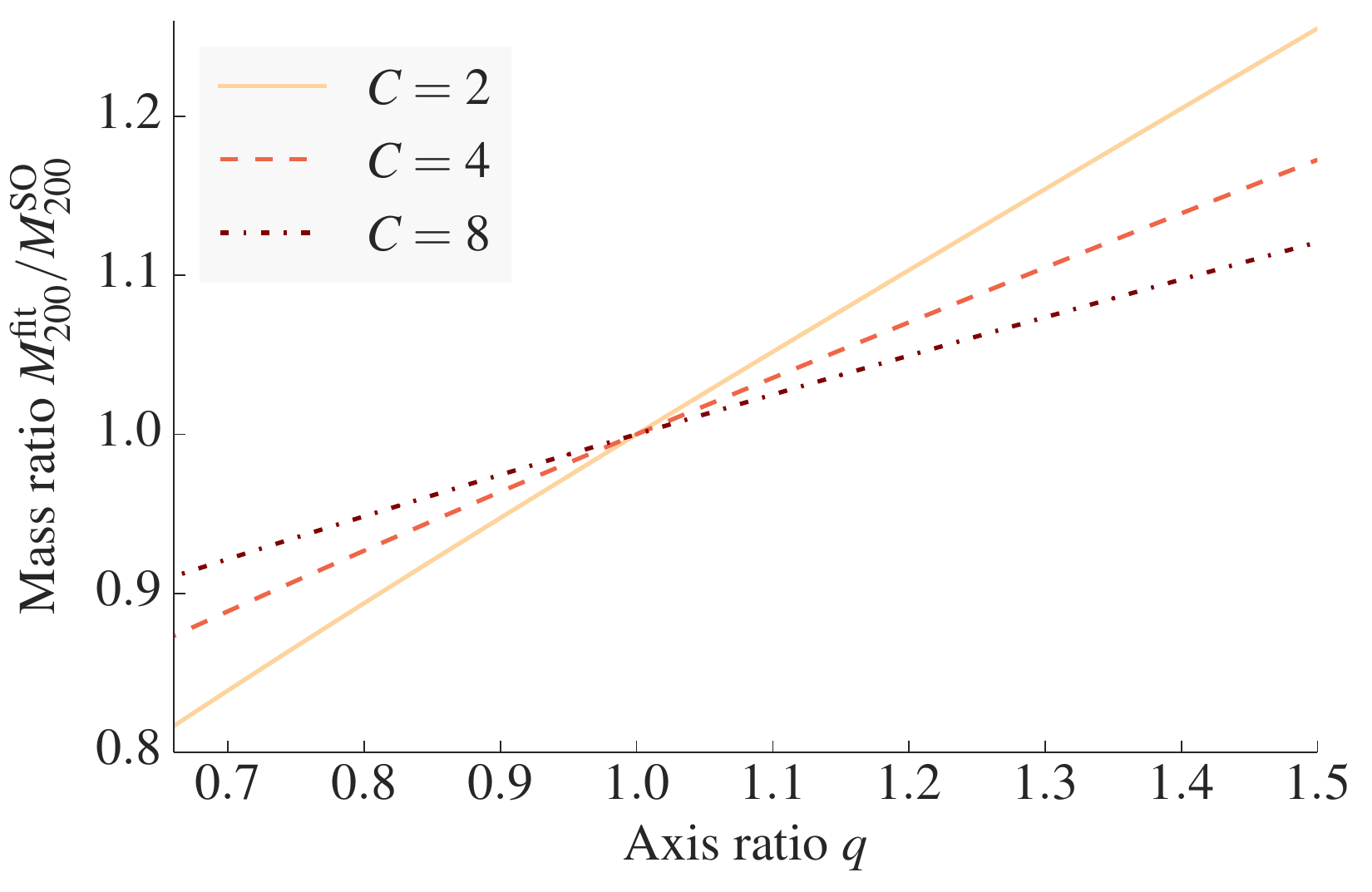}}
  \caption{The ratio of best fitting $M_{200}$ of a spherical NFW halo
    to the true spherical overdensity mass of an elliptical NFW
    halo. Different curves correspond to different concentration
    parameters.}
  \label{fig:mratio}
\end{figure}

We can find the spherical overdensity mass $M_{200}^\mathrm{SO}$ by
numerically solving
\begin{equation}
  \label{eq:5}
  \frac{3}{4\pi R^3_{200}} \int_0^{R_{200}} \int_0^\pi \int_0^{2\pi} 
  \d\phi\,\d\theta\,\d r r^2 \sin(\theta) \rho(\xi) = 200 \rho_\mathrm{crit}
\end{equation}
for $R_{200}$.  In Fig.~\ref{fig:CK2NFW} we show the ratio of SO
masses to \citetalias{2007MNRAS.380..149C} masses for three different
NFW concentration parameters. These two mass definitions agree within
1\%--2\% for a wide range of realistic axis ratios
\citep{2005ApJ...629..781K}. In the same figure we also show the mass
ratio for the best fit spherical NFW profile. Following the procedure
outlined by \citetalias{2007MNRAS.380..149C}, we fitted a spherical
NFW with free concentration and $r_{200}$ to the average density on
spherical shells. The result we find differs significantly from that
reported by \citetalias{2007MNRAS.380..149C}, who find differences of
up to $\sim30\%$ between this NFW mass the
\citetalias{2007MNRAS.380..149C} mass for the axis ratios studied
here. We are thus not able to reproduce the result
\citetalias{2007MNRAS.380..149C} present in their appendix but
emphasise that -- using their mass definition -- we reproduce their
predictions of the mass bias incurred when elliptical haloes are fitted
under the assumption of sphericity.\footnote{The conversion between SO
  and \citetalias{2007MNRAS.380..149C} masses based on fitting
  isodensity shells to an NFW profile was not used anywhere else in
  \citetalias{2007MNRAS.380..149C} and -- as far as we can tell --
  also not in any papers citing \citetalias{2007MNRAS.380..149C}.}
Although our mass definition is very close to the one of
\citetalias{2007MNRAS.380..149C}, we show the mass bias of fitting
spherical NFW haloes to elliptical profiles in Fig.~\ref{fig:mratio}
for completeness and later reference.

\subsection{Spherical Inversion}
\label{sec:spherical-inversion}

\citetalias{2007MNRAS.380..149C} studied the impact of halo
ellipticity on the mass estimates obtained by fitting spherical NFW
haloes to the observed reduced shear. An alternative way to obtain
cluster mass estimates is the spherical inversion of the projected
mass inside a cylinder using the Abel transform \citep{Abel1826}, such
as described by \citet{2007ApJ...656...27J}:
\begin{equation}
  \label{eq:6}
  \rho^\mathrm{inv}(r) = \frac{1}{\pi} \int_r^\infty
  \frac{-\Sigma^\prime(R)}{\sqrt{R^2 - r^2}} \d R\;,
\end{equation}
where the prime denotes a derivative with respect to the radial
coordinate.

The surface mass density $\Sigma$ of any halo is obtained by
integrating the 3-dimensional density profile along the LOS,
\begin{equation}
  \label{eq:7}
  \begin{split}
    \Sigma^\mathrm{ell}(r) & = 2 \int_0^\infty \rho[\xi(r, \theta, z)] \d z\\
    & = 2 \int_{qr}^\infty \rho(\xi) \frac{\xi}{\sqrt{\xi^2 - q^2r^2}} \d \xi
  \end{split}
\end{equation}
Recalling that for a spherical profile (the inversion of eq.~(\ref{eq:6}))
\begin{equation}
  \label{eq:8}
  \Sigma^\mathrm{sph}(r) = 2\int_r^\infty \rho(R) 
  \frac{R}{\sqrt{R^2 - r^2}} \d R\;,
\end{equation}
we see that
\begin{equation}
  \label{eq:9}
  \Sigma^\mathrm{ell}(r) = \Sigma^\mathrm{sph}(qr)\;.
\end{equation}
This means that there is an analytical degeneracy between the surface
mass density of spherical density profile and the surface mass density
of an elliptical density profile with its odd axis aligned with the
LOS. The two can be transformed into each other by a simple
rescaling. Consequentially weak lensing alone, or any other method
which depends linearly on the density, is unable to recover
3-dimensional mass information. For example, an elliptical NFW profile
is exactly degenerate with a spherical NFW profile of different mass
and concentration. Although this result is very straightforward to
derive, it seems to us that it is not widely known. Recently
\citet{2012MNRAS.425.2169G} reported their finding that the projected
density profile shapes seem to be independent from the underlying
3-dimensional density distribution in a numerical study of massive
galaxy clusters. This is easily explained by the degeneracy we just
described.

With this degeneracy, the inversion in eq.~(\ref{eq:6}) of an
elliptical profile then of course recovers a rescaled spherical
profile. As an illustration we consider the generalised NFW (gNFW)
profile
\begin{equation}
  \label{eq:10}
    \rho_\mathrm{gNFW}(r; r_\mathrm{s}) =
    \frac{\delta_\mathrm{c}\rho_\mathrm{c}}{(r/r_\mathrm{s})^\alpha
    \left(1 + r/r_\mathrm{s}\right)^\beta}\;,
\end{equation}
with scale radius $r_\mathrm{s}$. The standard NFW profile is
recovered for $(\alpha, \beta) = (1, 2)$ and a singular isothermal
sphere is obtained from $(\alpha, \beta) = (2, 0)$. We show in
Appendix~\ref{sec:abel-inversion-nfw} that if an elliptical radius
$\xi$ is used in eq.~(\ref{eq:10}) instead of the spherical radius $r$,
the Abel inversion leads to a density profile of
\begin{equation}
  \label{eq:11}
  \rho_\mathrm{gNFW}^\mathrm{inv}(r; q, r_\mathrm{s}) = q
  \rho_\mathrm{gNFW}(r; r_\mathrm{s} / q)\;.
\end{equation}

\section{Methods}
\label{sec:ellipticity-n-body}

Before we proceed to measure the ellipticity introduced into stacks of
optical selected clusters as a consequence of cluster finding
orientation bias and richness estimation orientation bias, we describe
our data sets and methods. Since knowledge of the true halo
ellipticity is required to quantify the size of the orientation bias,
we must work on simulations where this information is readily
available. These simulations must be realistic enough to run optical
cluster finders on them and get results that resemble reality. We
describe such a set of simulations as well as a suite of cluster
finders applied to them in following subsections before defining how
we measure the ellipticity of haloes and clusters.

\subsection{Simulations}
\label{sec:simulations}
For this study we have used the mock galaxy catalogues created for the
Dark Energy Survey based on the algorithm Adding Density Determined
GAlaxies to Lightcone Simulations (ADDGALS; Wechsler et al. in prep.;
Busha et al., in prep.). These are the same catalogs used by
\citet{2014MNRAS.440.2191S}. We reproduce our description of these
simulations from that paper in the following for the convenience of
the reader.

The ADDGALS algorithm attaches synthetic galaxies, including multiband
photometry, to dark matter particles in a lightcone output from a dark
matter $N$-body simulation and is designed to match the luminosities,
colours, and clustering properties of galaxies. The catalogue used
here was based on a single “Carmen” simulation run as part of the
LasDamas of simulations (McBride et al, in
preparation)\footnote{Further details regarding the simulations can be
  found at
  \texttt{http://lss.phy.vanderbilt.edu/lasdamas/simulations.html}}. This
simulation modeled a flat $\Lambda$CDM universe with
$\Omega_\mathrm{m} = 0.25$ and $\sigma_8 = 0.8$ in a 1 Gpc/$h$ box
with 11203 particles. A 220 sq deg light cone extending out to $z =
1.33$ was created by pasting together 40 snapshot outputs.  The galaxy
population for this mock catalogue was created by first using an
input luminosity function to generate a list of galaxies, and then
adding the galaxies to the dark matter simulation using an empirically
measured relationship between a galaxy's magnitude, redshift, and
local dark matter density, $P(\delta_\mathrm{dm}|M_\mathrm{r}, z)$ –
the probability that a galaxy with magnitude $M_\mathrm{r}$ and
redshift $z$ resides in a region with local density
$\delta_\mathrm{dm}$. This relation was tuned using a high resolution
simulation combined with the SubHalo Abundance Matching technique that
has been shown to reproduce the observed galaxy 2-point function to
high accuracy
\citep{2004ApJ...609...35K,2006ApJ...647..201C,2013ApJ...771...30R}.

For the galaxy assignment algorithm, a luminosity function that is
similar to the SDSS luminosity function as measured in
\citet{2003ApJ...592..819B} is chosen, but evolved in such a way as to
reproduce the higher redshift observations (e.g., SDSS-Stripe 82,
AGES, GAMA, NDWFS and DEEP2). In particular, $\phi^*$ and $M$ are
varied as a function of redshift in accordance with the recent results
from GAMA \citep{2012MNRAS.420.1239L}.  Once the galaxy positions have
been assigned, photometric properties are added. Here, a training set
of spectroscopic galaxies taken from SDSS DR5 was used. For each
galaxy in both the training set and simulation $\Delta_5$, the
distance to the 5th nearest galaxy on the sky in a redshift bin, is
measured. Each simulated galaxy is then assigned a spectral energy
distribution based on drawing a random training-set galaxy with the
appropriate magnitude and local density, k-correcting to the
appropriate redshift, and projecting onto the desired filters.  When
doing the colour assignment, the likelihood of assigning a red or a
blue galaxy is smoothly varied as a function of redshift in order to
simultaneously reproduce the observed red fraction at low and high
redshifts as observed in SDSS and DEEP2.  Haloes in the simulation are
identified by the \textsc{rockstar} phase-space halo finder
\citep*{2013ApJ...762..109B}.

Photometric noise and error estimates are added to the galaxy
catalogue based on the depth expected for the Dark Energy
Survey\footnote{\url{http://www.darkenergysurvey.org/}} (DES),
corresponding to $5\sigma$ detection limits of $\{26.0, 25.5, 24.8,
  24.3, 22.5\}$\,mag in grizY bands, respectively. This results in a
  total number of about 21 million galaxies extending out to redshift
  1.35. 

\subsection{Cluster Finders}
\label{sec:cluster-finders}

A plethora of galaxy cluster finders has been developed since the
proposal of \citet{1996AJ....111..615P} to use a spatial matched
filter algorithm on single passband data. Improvements in methodology
have come primarily from the inclusion of multi-band photometry
\citep[e.g.][]{2000AJ....120.2148G}, which can either be used as a
multi-dimensional colour-space in which to identify
overdensities \citep[e.g.][]{2005AJ....130..968M} or for estimating
photometric redshifts
\citep[e.g.][]{2010MNRAS.406..673M}. Improvements to the spatial
filtering have also been proposed, often in the form of Voronoi
tessellations \citep[e.g.][]{2002AJ....123...20K,2011ApJ...727...45S}.

We ran a total of four different cluster finders on the mock catalogues
generated as described in the previous section. The aim here is to
roughly cover the available space of modern cluster finder methods and
study whether different algorithms have different orientation biases
when finding galaxy clusters. We briefly describe each cluster finder
below. For all clusters found with these different methods an estimate of
their optical richness was computed using the $\lambda$ richness
estimator of \citet{2012ApJ...746..178R}.

\subsubsection{\textsc{redMaPPer}}
\label{sec:redmapper}
\textsc{redMaPPer} \citep{2014ApJ...785..104R} is a photometric
cluster algorithm that identifies galaxy clusters as over-densities of
red-sequence galaxies. The algorithm is divided into two stages: a
calibration stage and a cluster-finding stage. In the calibration
phase, redMaPPer empirically determines the colour distribution (mean
and scatter) of red-sequence galaxies as a function of redshift and
magnitude. This is achieved with an iterative procedure: using an
a-priori red-sequence model, seed galaxies with spectroscopic
redshifts in clusters are grouped with nearby potential cluster
members based on colour, which are then used to calibrate the
red-sequence. This model is used to re-estimate membership for every
galaxy, and the red-sequence model is then re-estimated. The procedure
is iterated until convergence. Once calibration is achieved, cluster
finding is performed. All galaxies are considered candidate cluster
centers, and assigned a redshift using our red-sequence model. Using
this redshift, cluster members are found, and a new cluster redshift
is estimated by simultaneously fitting the cluster members. The
procedure is iterated until convergence, and then the cluster is
recentered on the best possible central galaxy.  The list of clusters
is then rank-ordered and, in an iterative process called
\emph{percolation}, galaxies are probabilistically assigned to
clusters to ensure that no cluster is counted multiple times.

\subsubsection{\textsc{gmBCG}}
\label{sec:gmbcg}
Galaxy clusters almost always contain a brightest cluster galaxy (BCG)
at their centers and their member galaxies tend to cluster tightly in
the color space.  The \textsc{gmBCG} cluster
\citep{2010ApJS..191..254H} finder utilizes these two features to find
galaxy clusters.

Starting from a galaxy catalog, \textsc{gmBCG} first searches for BCG
candidates by applying a user-adjustable luminosity and color cut, and
then tries to model the color distribution of galaxies surrounding a
BCG candidate with a Error Corrected Gaussian Mixture Model
\citep{2009ApJ...702..745H}. If the final model contains a very narrow
red Gaussian component that corresponds to cluster red sequence
galaxies, as well as a wider and bluer Gaussian component that
corresponds to projected foreground/background galaxies and cluster
``blue cloud'' galaxies, \textsc{gmBCG} will claim to have found a
cluster candidate, and counts the galaxies falling into the red
sequence gaussian component as the candidate's member
galaxies. Finally, gmBCG ranks all the cluster candidates by their
member galaxy number counts and purges candidates according to their
members' spatial distribution and whether or not they can be included
in a more massive nearby cluster.

\subsubsection{\textsc{C4}}
\label{sec:c4}
The C4 algorithm identifies galaxies that exist in significant colour
overdensities (compared to a model colour-volume density generated
from random locations within the survey), and then groups them into
clusters. It does so without making assumptions about the combined
colour distribution of galaxy populations within clusters. The
original C4 algorithm relied on complete spectroscopic redshift
information \citep{2005AJ....130..968M}. The version used herein has
been adapted to surveys where only photometric redshift information is
available (Baruah et al. in prep).  Distances out to the 6th nearest
neighbour are calculated in the celestial sphere for each C4
galaxy. Treating these distances as an inverse proxy for density, they
are used to define the candidate cluster centres. Iterating through
the candidate centre list, C4 galaxies are associated to a candidate
centre (become cluster members) if they (i) lie within a $50$\,Mpc
bin along the line of sight, and (ii) the surface-number density of
the cluster exceeds some threshold above the surface density of C4
galaxies in this redshift bin. The central regions for these clusters
are then defined by a radius that envelopes 25\% of each cluster's
membership.  The cluster candidates are then merged if the central
galaxy of a C4 cluster candidate can be found in the central region of
a larger C4 cluster candidate, and if centres are within $\pm 0.06
(1 + z)$ of one another. These clusters form the C4 cluster catalog
and are ranked in order of the local number density (calculated with
the inverse 6th neighbour distance, as above) of the central C4
cluster galaxy.

\subsubsection{\textsc{WAzP}}
\label{sec:wazp}
\textsc{WAzP} (Wavelet Adapted $z$ Photometric, Benoist et al.,
in prep.) is an optical cluster finder based on the identification of
galaxy overdensities in (right ascension, declination, photometric
redshift $z_\mathrm{phot}$) space. The underlying algorithm uses 2-d
(right ascension, declination) and 1-d ($z_\mathrm{phot}$) density
field reconstruction based on the wavelet transform following the
method proposed by \citet*{1998A&AS..127..335F}. The main steps of the
algorithm can be described as follows:

\begin{enumerate}
\item The galaxy catalogue is sliced along the photometric redshift
  axis in overlapping redshift bins of variable sizes in order to
  follow the evolution of the photometric redshift dispersion with
  redshift. In each slice galaxies are selected in some magnitude
  range around the expected $m^*(z)$, the characteristic magnitude of
  a cluster luminosity function (a Schechter function) at a redshift
  $z$.

\item Galaxies from each slice are used to reconstruct the projected
  galaxy density field based on a wavelet transform method. In this
  reconstruction, only scales likely corresponding to clusters are
  kept, by default between $\sim0.5$--$3$\,Mpc.

\item In each slice, peaks of the density field are extracted with
  SExtractor configured to run without any background and with an
  absolute threshold set to some galaxy number density per Mpc$^2$.

\item As clusters may propagate across several slices, peaks from
  consecutive slices are associated, leading to the construction of
  cylinders (right ascension, declination,
  $z_{\mathrm{phot-min}}-z_{\mathrm{phot-max}}$), defining volumes
  potentially containing one or several clusters.

\item For each cylinder the 1-d density field along the photometric
  redshift axis is computed based again on a wavelet transform
  method. One or several photometric redshift peaks are identified
  leading to a refinement in position and size of the clusters.
\end{enumerate}

\subsection{Ellipticity Measurement}
\label{sec:ellipt-meas}

A number of ways exist to measure the ellipticity of dark matter
haloes\footnote{In the following text we will only talk about haloes,
  although the discussion of the method applies equally well to galaxy
  clusters in simulations where the 3-d position of galaxies are
  known.}, e.g., see \citet{2012MNRAS.420.3303B} for a concise review
of methods based on the inertia tensor. We chose to use the iterative
reduced inertia tensor in this work. Briefly, we compute the reduced
tensor of the mass quadrupole moments
\begin{equation}
  \label{eq:12}
  \mathcal{M}  _{ij} = \sum_{p=1}^{N} m_p \frac{r_{p,i} r_{p,j}}{r_p^2}\;,
\end{equation}
for a halo with $N$ particles of mass $m_p$ at positions $\vec{r}_p =
(r_{p,1}, r_{p,2}, r_{p,3})^\mathrm{t}$ with respect to the halo
centre.  The eigenvalues of this tensor are the squares of the axis
lengths $(a, b, c)$ of an ellipsoid with the same mass quadrupole
moments as the galaxy distribution. From these the axis ratios $t = a
/ c$ and $u = b / c$ are computed. These define the initial elliptical
radius $R$ as in equation~(\ref{eq:1}). In further iteration steps the
numerator in eq.~(\ref{eq:12}) is replaced with
\begin{equation}
  \label{eq:13}
  \tilde{r}_p^2 = r_{p,1}^2 + \frac{r_{p,2}^2}{t^2} +
  \frac{r_{p,3}^2}{u^2}\;,
\end{equation}
and only particles with $\tilde{r}_p \leq R$ are included in the
recomputation of eq.~(\ref{eq:12}). This iteration is terminated when
after iteration $k$
\begin{equation}
  \label{eq:14}
  \left|1 - \frac{t_k}{t_{k-1}} \right| < 0.01 
  \qquad \mathrm{and} \qquad
  \left|1 - \frac{u_k}{u_{k-1}} \right| < 0.01\;.
\end{equation}
By construction the galaxy density in our simulation traces the Dark
Matter density and we use the galaxy distribution as proxy for the
Dark Matter distribution.  Instead of using the location of Dark
Matter particles in the $N$-body simulation, we employ the position of
galaxies in eqs.~(\ref{eq:12}) and (\ref{eq:13}) and $m_p = 1$. As we
average many haloes the sampling noise introduced by using galaxies
instead of the much more numerous Dark Matter particles is negligible
and our problem becomes computationally much more manageable.

For every halo, a central galaxy is defined and all the galaxies
within a $3$\,Mpc sphere around the halo center are extracted. We
stack these galaxies of different haloes according to binning of halo
mass or optical richness, and run the above iteration on stacks of
halo galaxies.

\section{Results}
\label{sec:results}

We now present the results of our ellipticity measurements of stacked
clusters. In our analysis we only consider clusters that have a clear,
unique best halo match, as described as follows. These matches are
described as two-way matches using proximity matching.  Proximity
matching operates by iterating through a list of haloes, from most to
least massive, and imposing the condition that the highest ranked
cluster (where each cluster ranking system is a proxy ordering for
mass, as determined by the cluster finder's own mass-ranking
mechanism) is matched within a redshift cylinder of $\Delta z =
\pm0.1$ and $1$\,Mpc radius local to the halo (halo-to-cluster
matching). We chose $10^{13}\,\mathrm{M}_\odot$ as the cut-off mass
for haloes in this process. Similarly, the clusters are matched to the
haloes within redshift cylinders, going through the clusters from
highest ranked to least highest ranked (cluster-to-halo matching). A
uniqueness constraint is also applied, such that if a cluster has been
matched to a halo previously, or vice versa (i.e. a match to a higher
mass/ranked object has been previously made), it is no longer an
eligible match for lower mass/ranked objects. Where a given cluster
matches to a halo, and the same halo matches to that cluster, it is
considered a unique, two-way match. This allows us to study the impact
of orientation bias on cluster selection as a function of cluster mass
without having to resort to observational proxies for mass, which in
turn might be subject to orientation bias themselves. In all our
analyses we take advantage of our knowledge of the true halo centres
by stacking on those rather than on the cluster centroids identified
by the cluster finders. We discuss this choice in
Sect.~\ref{sec:discussion}.

\begin{figure}
  \resizebox{\hsize}{!}{\includegraphics[clip]{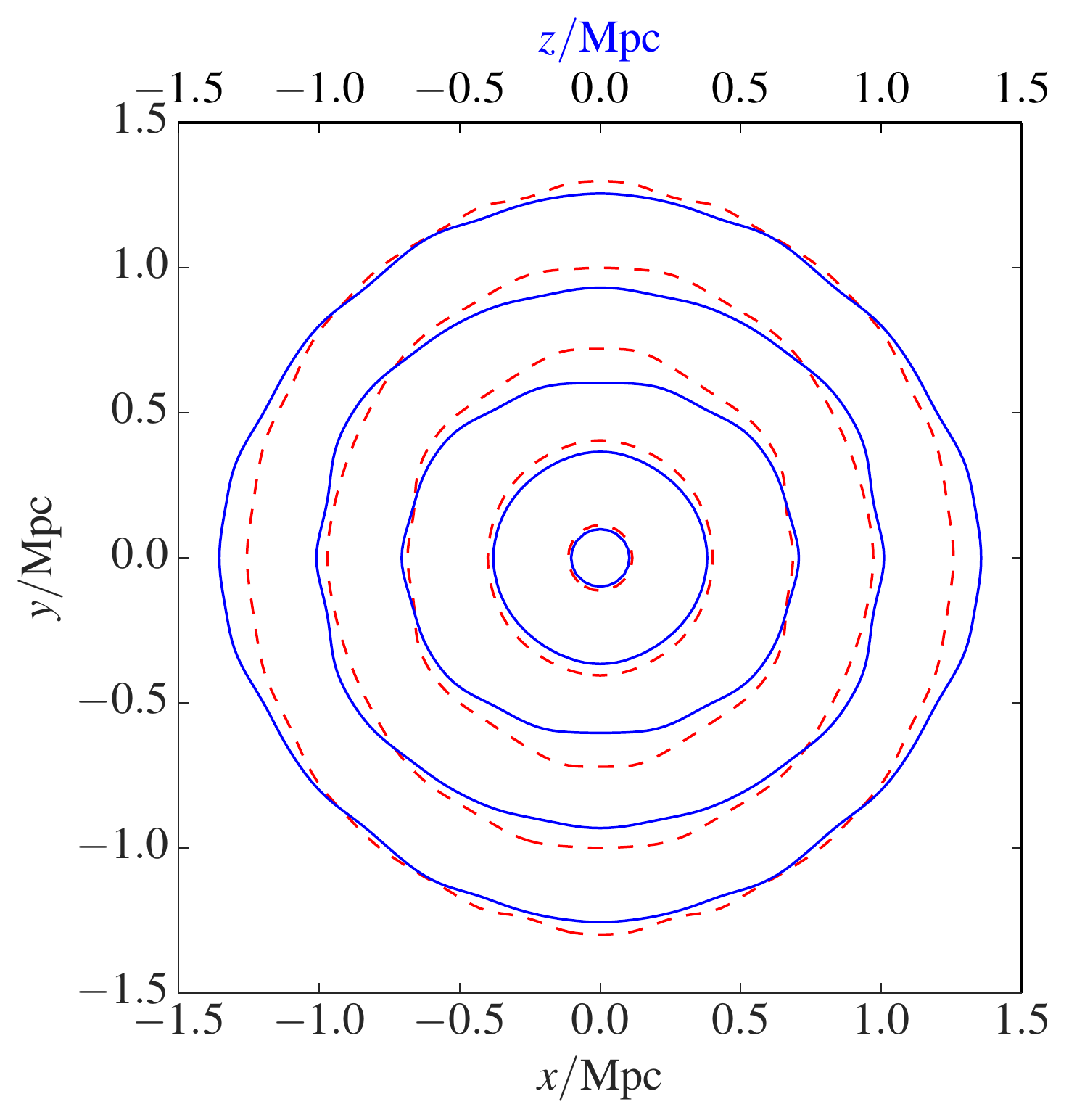}}
  \caption{Isodensity contours for the stack of all \textsc{redMaPPer}
    clusters for two different cross-sections through the stack. The
    clusters are stacked on the centres of the haloes to which they are
    matched. Red dashed contours are line of equal surface mass
    density in the $x$-$y$ plane (the plane of the sky). Solid blue
    contours are at the same density levels in a plane perpendicular
    to the sky (the $y$-$z$ plane). In this case the line-of-sight
    runs horizontally through the centre of the figure. To increase
    the signal-to-noise, these contours were generated by transforming
    the positions of all galaxies to the first quadrant and then
    reproducing that quadrant three times.}
  \label{fig:cluster_sections}
\end{figure}

To illustrate the reality of orientation bias when selecting
galaxy clusters, we produced a stack of all clusters found by the
\textsc{redMaPPeR} algorithm in the mock catalogues. In
Figure~\ref{fig:cluster_sections} we show isodensity contours for two
different cross-sections through the cluster stack.  On the one hand
the contours in the plane of the sky are circularly symmetric,
consistent with our expectation that cluster finders do not have a
preferred direction in this plane. On the other hand the isodensity
contours show a clear elongation along the $z$-axis, the LOS in our
choice of coordinate system. This establishes that orientation bias
exists at least for some sub-population of all clusters when they are
optically selected.

\begin{figure}
  \resizebox{\hsize}{!}{\includegraphics[clip]{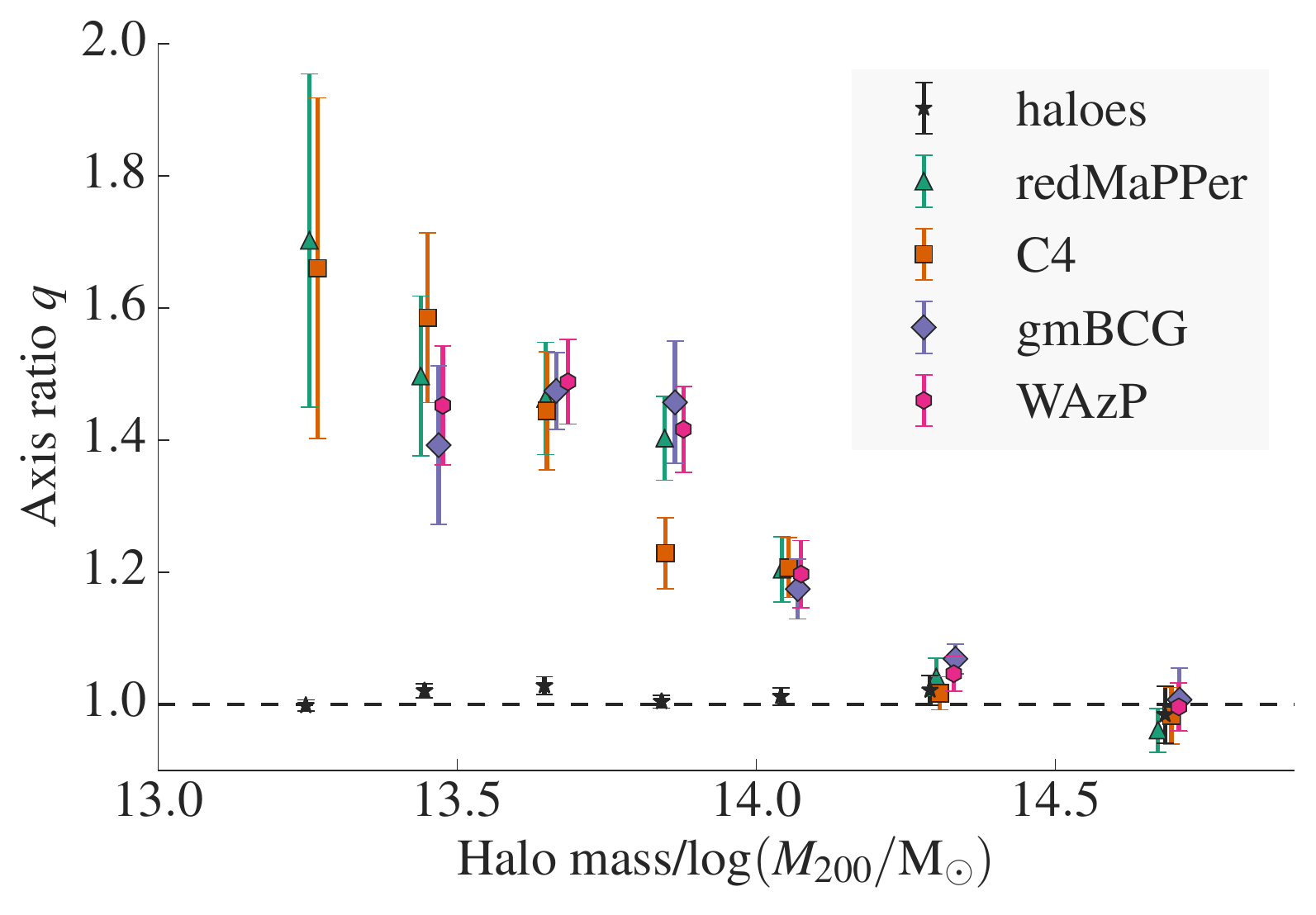}}
  \caption{Measured axis ratio $q$ as function of halo and cluster
    mass. Shown are only those clusters that could be uniquely matched
    to haloes. Error bars are computed from bootstrap resampling the
    haloes/clusters in a mass bin. Due to a minimum richness imposed
    by every cluster finder for inclusion in the cluster catalog fewer
    halos could be matched to clusters at low masses. Therefore the
    error bars of clusters at lower halo masses are larger than at
    high mass. The size of the error bars of the haloes has the
    opposite behaviour because more low mass haloes than high mass
    haloes exist and all haloes including those not matched to
    clusters are included in the analysis as a null test. A small
    intrinsic scatter between different cluster finders is present in
    each mass bin but the displayed points are offset in $x$-direction
    for clarity by an amount larger than this.}
\label{fig:m200-q}
\end{figure}

The average ellipticity of clusters binned by true cluster mass is
shown in Fig.~\ref{fig:m200-q}. At low masses galaxy clusters are
strongly prolate with axis ratios $\gtrsim 1.5$. This elongation along
the LOS decreases as the cluster mass exceeds
$\log(M_{200}/\mathrm{M}_\odot) \approx 14.1$ or about
$1.3\times10^{14}$\,M$_\odot$. At the highest masses, the stacks of
optically selected clusters are spherically symmetric. This trend is
true for all of the cluster finding algorithms studied
here. Figure~\ref{fig:m200-q} also shows the axis ratio for stacks of
haloes found in the $N$-body simulations used in the mocks. As
expected, these are consistent with being spherical.

We interpret this behaviour of the cluster finders as increasing
difficulty in identifying galaxy clusters at decreasing masses. A
$10^{15}$\,M$_\odot$ cluster is such an obvious overdensity that any
cluster finder will see it, regardless of its orientation. At lower
masses, finding galaxy clusters becomes more of a challenge and the
mechanism of orientation bias as described earlier in this paper
becomes effective. It is worth pointing out that this difficulty in
finding clusters depends on the intrinsic scatter in the mass-richness
relation. For a higher scatter in optical richness at fixed mass, the
probability of missing higher mass clusters increases because they may
have fewer galaxies. \citet{2012ApJ...746..178R} and
\citet{2014ApJ...783...80R} demonstrated that their $\lambda$ richness
estimator, which forms the basis of the \textsc{redMaPPeR} cluster
finder, has a comparatively low scatter in mass at fixed richness,
$\sigma_{\ln M|\lambda} \approx 0.2\text{--}0.3$, for an X-ray
selected cluster sample. A similarly low scatter $\sigma_{\ln M |
  \lambda}\sim 0.2 $ is also observed for a sample of $>200$ Planck
Sunyaev-Zeldovich (SZ) selected clusters
\citep{2014arXiv1401.7716R}. This is lower than the scatter in halo
mass at fixed $\lambda$ in our mock catalogues, which we find to be as
high as $0.8$. We therefore assume that the mocks have an intrinsic
scatter in the mass-richness relation that exceeds the scatter in the
real Universe. The effect for our study would be that the points in
Fig.~\ref{fig:m200-q} are shifted further to the right than they would
be in real data, i.e., the effect of orientation bias would be
overestimated in the mocks. In the absence of mock catalogues, which
reproduce the estimated real scatter in the mass-richness relation, we
tested the hypotheses that this scatter moves the location of data
points in Fig.~\ref{fig:m200-q} from left to right with a simulation
that has an even higher scatter using only the \textsc{redMaPPeR}
cluster finder. We found that indeed the points move further to
the right. It is difficult to estimate how much the curve in
Fig.~\ref{fig:m200-q} would shift to the left if $\sigma_{\ln
  M|\lambda}$ is indeed as low for optically selected clusters as
indicated by \citet{2012ApJ...746..178R} because many cluster finders'
completeness has a complex dependency on this quantity.

\begin{figure}
  \resizebox{\hsize}{!}{\includegraphics[clip]{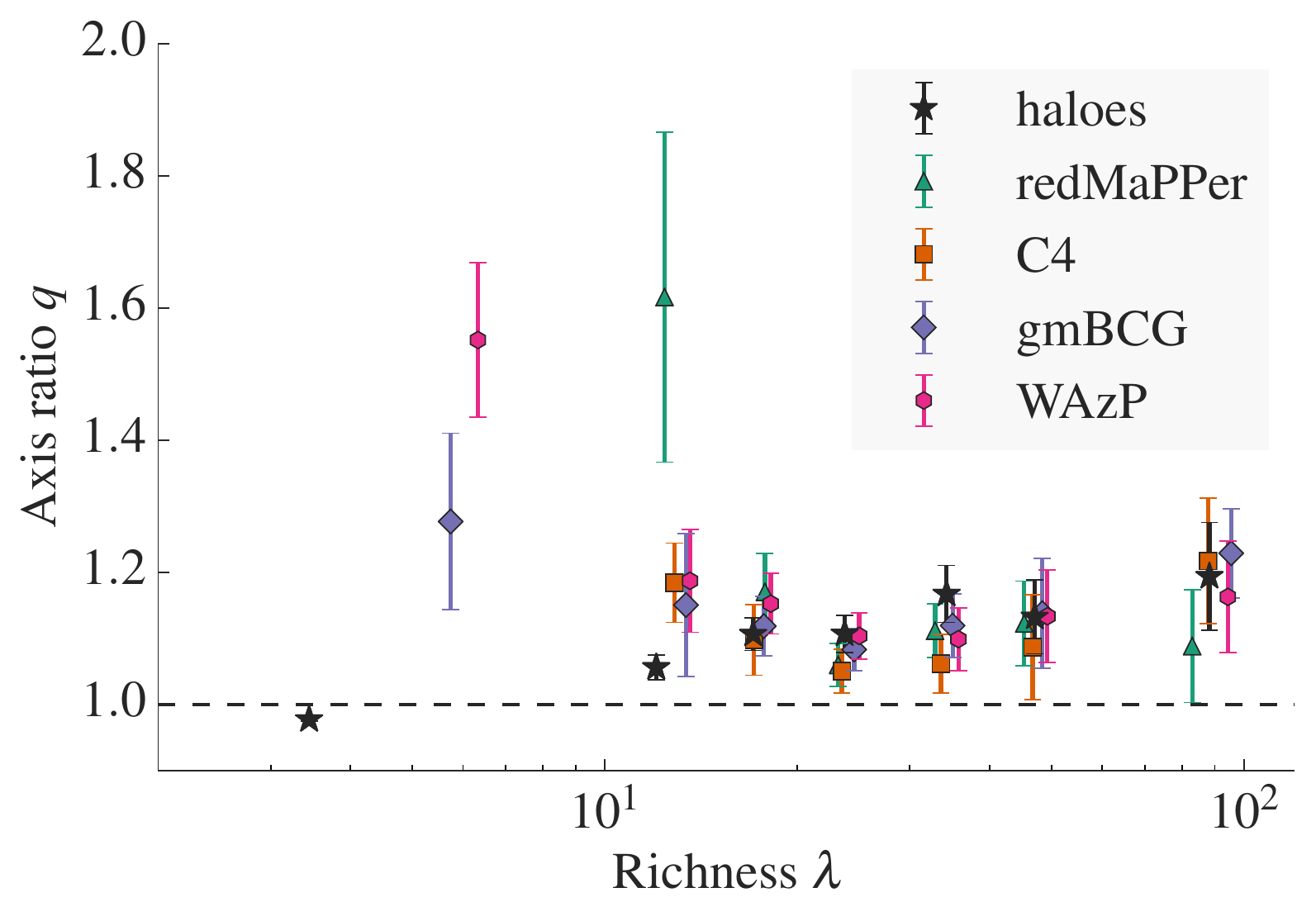}}
  \caption{Measured axis ratio $q$ as a function of estimated halo and
    cluster richness $\lambda$. Again horizontal offsets are applied
    to the data points for greater clarity.}
  \label{fig:ell_lambda}
\end{figure}

In real data the true mass of clusters is of course unknown and galaxy
clusters are binned by a proxy for their mass, often a richness
estimator in the case of optically selected clusters.
Figure~\ref{fig:ell_lambda} shows the measured axis ratio for haloes
and clusters binned by the optical richness estimator $\lambda$.
First, we notice that haloes, when binned by optical richness, are no
longer spherical. This confirms our initial hypothesis that
orientation bias is not only at work during the cluster finding
process but also during richness estimation. Prolate clusters appear
denser and thus richer on the sky. They are pushed to higher richness
bins, which then on average deviate from spherical symmetry. This
happens at the expense of the lowest richness bins, from which the
prolate haloes are removed. They then appear slightly oblate at the
lowest richness. The axis ratio of haloes increases with increasing
$\lambda$ until seemingly an approximate balance between prolate
haloes that are pushed up from lower richness bins and prolate haloes
that are pushed into the next higher redshift bin is established. This
seems to be the case at $\lambda \gtrsim 15$. We also note that the
haloes in the highest richness bin are marginally, but not
significantly, more prolate than in any other bins, further supporting
this scenario.

The behaviour is different for galaxy clusters. At $\lambda > 15$ the
measured axis ratios for all cluster finders are consistent with being
independent of richness and around $q \sim 1.1$. Significant outliers
occur at lower $\lambda$ for the \textsc{redMaPPer, gmBCG}, and
\textsc{WAzP} cluster finders. These values should be excluded from
any interpretation of the present study. The performance of
\textsc{redMaPPer} is only characterised and well understood at
$\lambda > 20$ \citep{2014ApJ...785..104R}, which is the threshold
adopted by the developers for inclusion of objects in the cluster
catalogue. It is also important to note that \textsc{redMaPPer}
cluster detection and $\lambda$ richness estimation are strongly
intertwined and tuned to each other. Thus the two lowest richness
points of the \textsc{gmBCG} and \textsc{WAzP} cluster are only
presented for completeness and should not be interpreted as having
reliable richness measurements and they should not be understood as an
ability of these cluster finders to find lower richness clusters than
\textsc{redMaPPer}. Given the complex interplay between cluster
selection and richness estimation, we make no attempt at an
interpretation of these points.

Binning by richness as in Fig.~\ref{fig:ell_lambda} does not reproduce
the trend of decreasing prolateness with increasing mass and thus
increasing richness seen in Fig.~\ref{fig:m200-q}.  We observe that
the scatter in optical richness leads to substantial mixing of cluster
masses between richness bins. We must, however, caution that there are
indications that the intrinsic scatter in the simulations exceeds that
of real data and thus artificially enhances this mixing.

Furthermore, the net effect of orientation bias to push clusters into
higher richness bins counteracts the decrease of axis ratios with
increasing mass.  The reason is that lower mass clusters, which are
more subject to orientation bias in the cluster finding step and thus
appear more elliptical, are preferentially measured to have higher
richness as compared to the seemingly rounder high mass clusters.
Thus low mass clusters are preferentially pushed into higher richness
bins, resulting in a higher measured mean ellipticity. The result is
that orientation bias acts as an additional correlated scatter between
cluster mass and richness at fixed mass.

\section{Summary \& Discussion}
\label{sec:discussion}

We have established that optical cluster selection and richness
estimation are subject to a bias heretofore unconsidered in the study
of optical cluster selection. Prolate galaxy clusters are found
preferentially as compared to spherical clusters, and their richness
is over-estimated.

We ran a wide variety of cluster finders to test the orientation bias
when selecting clusters. As a function of mass, all cluster finders
studied here show a similar orientation bias. The large scatters
associated with our axis ratio measurements also smear out any
possible difference in the behavior of different cluster finders. A
consequence of this orientation bias is that stacked weak-lensing
analyses of galaxy clusters violate the previously made assumption
that averaging over enough clusters makes the stacks spherically
symmetric. We find instead that binning optically selected galaxy
clusters by optical richness makes these stacks elliptical with axis
ratios of major over minor axes $q \sim 1.1$. The exact value and its
behaviour with richness likely depends on how much additional scatter
the richness estimator at fixed mass has, as well as the intrinsic
scatter of the mass--richness relation. The latter is larger in the
simulations we used than is expected in the real Universe and the
value of $q$ we find here is an upper limit when clusters can be
uniquely associated with haloes. For simplicity and because of its
reported low intrinsic scatter, we have tested only the $\lambda$
richness estimator \citep{2012ApJ...746..178R}. The similar
orientation bias of cluster finders when rank ordered by mass turns
into a similar orientation bias when clusters are rank ordered by the
same richness estimator.

The choice of stacking matched clusters instead of all clusters,
including false positives and clusters encompassing more than one
halo, was made to avoid miscentring. Optically selected clusters have
a certain rate of misidentified central galaxies, which serve as proxy
for the halo centre. If we were to stack on optically identified
clusters we would incur offsets not only in the plane of the sky but
also along the LOS as we know the 3-d position of galaxies and use it
to to select galaxies contributing to the computation of the inertia
tensor. Lensing on the other hand is not sensitive to such tiny
differences in redshift and miscentring along the $z$-axis. Because we
want to study the impact of orientation bias on lensing analyses of
optically selected clusters, we chose to avoid these complications
caused by miscentring by limiting the analysis to matched
clusters. However, the unmatched clusters contain a certain fraction
of false positives, structures that do not correspond to a cluster
size halo or the superposition of two close clusters. These would
typically boost the observed ellipticity and this enhanced ellipticity
would also bias the lensing signal. We find that stacking all clusters
indeed leads to somewhat but not significantly higher $q$ values. We
made no attempt to determine whether this difference is primarily
caused by elongated structures erroneously selected as clusters, or by
miscentring along the $z$-axis. We emphasize that these higher $q$
values cannot be directly propagated into lensing mass biases since
they are affected by miscentring along the LOS, which does not impact
weak-lensing mass estimates. We did not address the question whether
such a bijectively matched sub-sample of galaxy clusters used in this
work could be identified in survey data when additional observables
such as velocity dispersions, X-ray morphology, location on scaling
relations, etc. are available. These questions can be addressed with
improved simulations and future large multi-wavelength surveys.

It is well established that ellipticities in clusters lead to biases
in cluster mass estimation \citepalias{2007MNRAS.380..149C}. We find
that the size of this effect for the spherical overdensity mass
definition, which is the basis for the commonly used
\citet{2008ApJ...688..709T} cluster mass function, is very similar to
that of the cluster mass definition of
\citetalias{2007MNRAS.380..149C}. We emphasize that an analytical
degeneracy exists for the projected density profiles of spherical
haloes with elliptical haloes of a different mass and
concentration. Weak lensing alone is thus unable to determine the
magnitude of the orientation bias and its resultant bias in cluster
mass calibration. For the axis ratios of $q \sim 1.1$ we find a mass
bias of $3\text{--}6$ per cent -- depending on cluster concentration
-- is expected from Fig.~\ref{fig:mratio}.

Misestimation of cluster masses will contribute to inconsistencies
between galaxy cluster scaling relations derived from different
observables. A prominent example of such a discrepancy is the mismatch
between the observed integrated Sunyaev-Zeldovich signal
$Y_\mathrm{SZ}$ in early Planck data and the one predicted from
cluster scaling relations \citep{2011A&A...536A..12P}. In this case,
optical richness $N_{200}$ was related to cluster mass
\citep{2007arXiv0709.1159J,2009ApJ...699..768R}, for which in turn
scaling relations for X-ray luminosity $L_\mathrm{X}$ and
$Y_\mathrm{SZ}$ were used to predict the integrated Compton-$y$ for
given optical richnesses. The predicted values were significantly
higher than the observed SZ signal. A similar effect has been observed
by the Atacama Cosmology Telescope \citep{2013ApJ...767...38S}.

Orientation bias can contribute to such discrepancies. The mass of
prolate cluster stacks is overestimated so that $M-N_{200}$ scaling
relations predict cluster masses at a given richness that are too
high. This in turn leads to higher $Y_\mathrm{SZ}$ values, as observed
by the \citet{2011A&A...536A..12P}. The magnitude of this affect, a
rescaling of the cluster masses by $3\text{--}6$ per cent, is not
enough to explain this particular discrepancy fully, so that other
effects like miscentring \citep{2012ApJ...757....1B} and
underestimated uncertainties in the X-ray scaling relations
\citep{2014MNRAS.438...49R,2014MNRAS.438...62R} are needed in this
case. \citet{2014MNRAS.438...78R} showed that a self-consistent
treatment of the scaling relations and proper inclusion of previously
unaccounted systematic errors can resolve the tension found by the
\citeauthor{2011A&A...536A..12P}. As part of this process
\citet{2014MNRAS.438...78R} lowered their weak-lensing
mass-calibration by $10$\,per cent, a correction they attribute to
intrinsic covariance between weak lensing mass and cluster richness at
fixed redshift. Orientation bias also induces such correlated scatter
and although not isolated in that analysis, it is implicitly included.

A similar orientation bias is known to exist in the SZ selection of
galaxy clusters \citep{1991ApJ...379..466B} and has mostly been
discussed in the context of measurement of the Hubble parameter using
the Sunyaev-Zel'dovich effect
\citep[e.g.][]{2005MNRAS.357..518J}. However, spatially unresolved
observations -- such as the Planck data -- see the total integrated
pressure.  No additional correlation between the ellipticity of the
optical \textsc{maxBCG} clusters and the measured $Y_\textsc{SZ}$ is
expected from an SZ orientation bias in the
\citep{2011A&A...536A..12P} results.

Weak-lensing mass-calibration biases are propagated into cosmological
parameter estimates, where scaling relations based on them are
used. Assessing the bias in the determination of cosmological
parameters caused by the overestimation of cluster masses due to
orientation bias in previous studies is not straightforward. The
additional correlated scatter caused by the orientation bias can lead
to complex parameter degeneracies. This is one reason we cannot simply
correct the cosmological parameters of \citet{2010ApJ...708..645R}
without re-running the entire MCMC. The other reason is that the their
optical richness estimator is different from the $\lambda$ richness
estimator used in this work. We can, however, make an approximate
determination of this bias by following \citet{2013arXiv1302.5086R}:
Low redshift clusters essentially constrain the quantity $s_8 =
\sigma_8 \left(\Omega_\mathrm{m}/0.25\right)^{\eta}$, where $\eta
\approx 0.4\text{--}0.5$. \citet{2009ApJ...692.1033V} showed that
shifting the masses of all galaxy clusters in their sample by $\pm9$
per cent shifts the $s_8$ value by $\pm0.024$. For small shifts
$\Delta \ln M_{200}$ we use a linear approximation
\begin{equation}
  \label{eq:20}
  s_8 = s_{8,0} + 0.024 \frac{\Delta \ln M_{200}}{0.09}\;.
\end{equation}
Under the assumption that the mass calibration bias is $6$ per cent,
the maximum value suggested by an axis ratio of $q=1.1$, the
corresponding shift in $s_8$ is $0.016$. This is a factor two smaller
than the error found by \citet{2010ApJ...708..645R}. Unless the
\textsc{maxBCG} cluster finder, for which they performed a
weak-lensing cluster mass calibration, or their $N_{200}$ richness
estimator perform significantly differently from any of the cluster
finders and the $\lambda$ richness estimator studied here, the
cosmological parameter constraints of \citet{2010ApJ...708..645R}
should not be significantly affected by orientation bias. Moreover,
\citet{2010ApJ...708..645R} considered the possibility that their
lensing masses of galaxy clusters are biased by introducing a bias
parameter $\beta$. Their best fit value is $\beta = 1.016 \pm 0.060$,
entirely consistent with the mass bias expected from orientation bias.

Extending such forecasts to upcoming surveys like DES is not
straightforward. DES is expected to find about 100\,000 optically
selected clusters out to a redshift $z \sim 1$. Probing the evolution
of the cluster abundance with redshift helps breaking the
$\Omega_\mathrm{m}\text{--}\sigma_8$ degeneracy. Biases in mass
calibration will then no longer move simply along the $s_8$
line. Forecasts made by \citet{2013JCAP...02..030K} for DES based on a
maxBCG \citep{2007ApJ...660..239K} like optical cluster selection
indicate that the marginalised $1\sigma$ error on $\Omega_\mathrm{m}$
should decrease by a factor of $2.5$ compared to
\citet{2010ApJ...708..645R} with a follow-up program obtaining cluster
masses with an accuracy of $30\%$ for only 100 clusters. Already at
this level the systematic error induced by orientation bias would
dominate over the statistical error. Such a limited follow-up is an
extremely pessimistic assumption since there are already more clusters
with observations from the South Pole Telescope
\citep{2013ApJ...763..127R} and DES can obtain weak-lensing mass
measurements via its own observations. In any case, the statistical
power of  surveys like the DES will be so substantial as to demand
improved  understanding of the effects of orientation bias.

\section*{Acknowledgements}
\label{sec:acknowledgements}
JPD thanks Virginia Corless for detailed discussions on
Sect.~\ref{sec:predictions}. This work was supported in part by NSF
grant AST-0807304, by U.S. Department of Energy grant
DE-FG02-95ER40899, and by the U.S. Department of Energy contract to
SLAC no. DE-AC02-76SF00515.

Funding for the DES Projects has been provided by the U.S. Department
of Energy, the U.S. National Science Foundation, the Ministry of
Science and Education of Spain, the Science and Technology Facilities
Council of the United Kingdom, the Higher Education Funding Council
for England, the National Center for Supercomputing Applications at
the University of Illinois at Urbana-Champaign, the Kavli Institute of
Cosmological Physics at the University of Chicago, Financiadora de
Estudos e Projetos, Funda\,c\~ao Carlos Chagas Filho de Amparo \`a
Pesquisa do Estado do Rio de Janeiro, Conselho Nacional de
Desenvolvimento Cient\'ifico e Tecnol\'ogico and the Minist\'erio da
Ci\^encia e Tecnologia, the Deutsche Forschungsgemeinschaft and the
Collaborating Institutions in the Dark Energy Survey.

The Collaborating Institutions are Argonne National Laboratories, the
University of California at Santa Cruz, the University of Cambridge,
Centro de Investigaciones Energeticas, Medioambientales y
Tecnologicas-Madrid, the University of Chicago, University College
London, the DES-Brazil Consortium, the Eidgen\"ossische Technische
Hochschule (ETH) Zurich, Fermi National Accelerator Laboratory, the
University of Edinburgh, the University of Illinois at
Urbana-Champaign, the Institut de Ciencies de l'Espai (IEEC/CSIC), the
Institut de Fisica d'Altes Energies, the Lawrence Berkeley National
Laboratory, the Ludwig-Maximilians Universit\"at and the associated
Excellence Cluster Universe, the University of Michigan, the National
Optical Astronomy Observatory, the University of Nottingham, the Ohio
State University, the University of Pennsylvania, the University of
Portsmouth, SLAC National Laboratory, Stanford University, the
University of Sussex, and Texas A\&M University.

This paper has gone through internal review by the DES collaboration.

\bibliography{library}
\appendix
\onecolumn
\section{Abel inversion of an elliptical gNFW halo}
\label{sec:abel-inversion-nfw}

Here we show that the spherical Abel inversion of a generalised
elliptical NFW halo leads to a density profile that is described by a
rescaling of the gNFW density~(\ref{eq:10}) and its scale radius. The
surface mass density is obtained by inserting eq.~(\ref{eq:10}) into
eq.~(\ref{eq:7}) and using our definition of the elliptical radius
$\xi^2 = q^2 r^2 + z^2$,  
\begin{equation}
  \label{eq:15}
  \Sigma(r; q, r_\mathrm{s}) = 2 \int_0^\infty \d  z
  \frac{\delta_\mathrm{c} \rho_\mathrm{c}}{\left(\sqrt{q^2 r^2 +
      z^2}/r_\mathrm{s}\right)^\alpha \left(1 + 
      \sqrt{q^2r^2 + z^2} / r_\mathrm{s}\right)^\beta} 
\end{equation}
For $(\alpha, \beta, q) = (1, 2, 1)$ this integral has been explicitly
evaluated by \citet{1996A&A...313..697B} and
\citet{2000ApJ...534...34W}. We instead choose a different route and
compute the derivative $\Sigma^\prime = \frac{\d \Sigma}{\d r}$ of
eq.~(\ref{eq:15}),
\begin{equation}
  \label{eq:16}
  \Sigma^\prime(r; q, r_\mathrm{s}) 
  = -2 q^2 r \int_{qr}^\infty \frac{\d \xi}{\xi \sqrt{\xi^2 -
      q^2r^2}} \frac{(\alpha+\beta)\xi + \alpha r_\mathrm{s}}{\xi +
    r_\mathrm{s}} \rho_\mathrm{gNFW}(\xi; r_\mathrm{s})
\end{equation}
The inverse Abel transform of this surface mass density profile is
obtained by substituting~(\ref{eq:16}) into~(\ref{eq:6}),
\begin{equation}
  \label{eq:17}
  \rho_\mathrm{gNFW}^\mathrm{inv}(\xi; q, r_\mathrm{s}) = \frac{2}{\pi} 
  \delta_\mathrm{c} \rho_\mathrm{c} r_\mathrm{s}^{\alpha+\beta}q^2
  \int_\xi^\infty \frac{r\, \d  r}{\sqrt{r^2 - \xi^2}} \int_{qr}^\infty
  \frac{\d  \xi}{\xi^{\alpha + 1} \sqrt{\xi^2 - q^2r^2}} \frac{(\alpha+\beta) \xi +
    \alpha r_\mathrm{s}}{(\xi + r_\mathrm{s})^{(\beta+1)}}\;.
\end{equation}
This may be rewritten in terms of a gNFW profile using the
substitutions $qu = \xi$ and $qu_\mathrm{s} = r_\mathrm{s}$,
\begin{equation}
  \label{eq:18}
  \begin{split}
    \rho_\mathrm{gNFW}^\mathrm{inv}(\xi; q, r_\mathrm{s}) &=
    \frac{2q \delta_\mathrm{c} \rho_\mathrm{c} u^{\alpha+\beta}_\mathrm{s}}{\pi}
    \int_\xi^\infty \frac{r\, \d  r}{\sqrt{r^2 - \xi^2}} 
    \int_r^\infty \frac{\d  u}{u^{\alpha+1} \sqrt{u^2 - r^2}} 
    \frac{(\alpha+\beta)u +\alpha u_\mathrm{s}}{(u + u_\mathrm{s})^{\beta+1}} \\
    &= \frac{2q}{\pi} \int_\xi^\infty \frac{r\, \d  r}{\sqrt{r^2 - \xi^2}}
    \int_r^\infty \frac{\d  u}{u \sqrt{u^2 - r^2}}
    \frac{(\alpha+\beta)u + \alpha u_\mathrm{s}}{u + u_\mathrm{s}}
    \rho_\mathrm{gNFW}(u; r_{s}/q) \; .
  \end{split}
\end{equation}
We see now that eq.~(\ref{eq:18}) is really the expression for a
spherical gNFW profile,
\begin{equation}
  \label{eq:19}
  \begin{split}
  \rho_\mathrm{gNFW}^\mathrm{inv}(\xi; q,r_\mathrm{s}) &= 
  \frac{-q}{\pi} \int_\xi^\infty \frac{\Sigma^\prime(r, q=1,
    u_\mathrm{s}=r_\mathrm{s}/q) 
    \d  r}{\sqrt{r^2 - \xi^2}} \\
  &= q \rho_\mathrm{gNFW} (\xi; r_\mathrm{s}/q) \; .
  \end{split}
\end{equation}

\bsp
\label{lastpage}
\end{document}

%% file: affiliations.tex
\newcommand{\afsep}{$^,$}
\newcommand{\Munich}{$^{1}$}
\newcommand{\ExcellenceCluster}{$^{2}$}
\newcommand{\AnnArborPhysics}{$^{3}$}
\newcommand{\KIPAC}{$^{4}$}
\newcommand{\Stanford}{$^{5}$}
\newcommand{\SLAC}{$^{6}$}
\newcommand{\AnnArborAstronomy}{$^{7}$}
\newcommand{\Sussex}{$^{8}$}
\newcommand{\OCA}{$^{9}$}
\newcommand{\LIneA}{$^{10}$}
\newcommand{\ON}{$^{11}$}
\newcommand{\FNAL}{$^{12}$}

%% file: abstract.txt
Weak-lensing measurements of the averaged shear profiles of galaxy
clusters binned by some proxy for cluster mass are commonly converted
to cluster mass estimates under the assumption that these cluster stacks
have spherical symmetry. 
In this paper we test whether this assumption holds for optically
selected clusters binned by estimated optical richness.
Using mock catalogues created from $N$-body simulations populated
realistically with galaxies, we ran a suite of optical cluster finders
and estimated their optical richness. We binned galaxy clusters by true
cluster mass and estimated optical richness and measure the
ellipticity of these stacks.
We find that the processes of optical cluster selection and richness
estimation are biased, leading to stacked structures that are
elongated along the line-of-sight. We show that weak-lensing alone
cannot measure the size of this orientation bias.
Weak lensing masses of stacked optically selected clusters are
overestimated by up to $3\text{--}6$\,per cent when clusters can be
uniquely associated with haloes. This effect is large enough to lead
to significant biases in the cosmological parameters derived from
large surveys like the Dark Energy Survey, if not calibrated via
simulations or fitted simultaneously. This bias probably also
contributes to the observed discrepancy between the observed and
predicted Sunyaev-Zel'dovich signal of optically-selected clusters.
